\documentclass[nobm,nocite]{epl}

\title{Characteristic features of the temperature dependence\\
of the surface impedance in polycrystalline \chem{Mg B_2} samples}
\author{Yu.A.~Nefyodov\inst{1} \and M.R.~Trunin\inst{1} \and A.F.~Shevchun\inst{1} \and
D.V.~Shovkun\inst{1} \and N.N.~Kolesnikov\inst{1} \and
M.P.~Kulakov\inst{1} \and A.~Agliolo Gallitto\inst{2} \and
S.~Fricano\inst{2}}
\institute{\inst{1}Institute of Solid State
Physics, 142432 Chernogolovka, Moscow district, Russia\\
\inst{2}INFM, Unita di Palermo and Dipartimento di Scienze Fisiche
e Astronomiche\\ Via Archirafi 36, I-90123 Palermo, Italy}

\pacs{74.25.Nf}{Response to electromagnetic fields (nuclear
magnetic resonance, surface impedance, etc.)  }
\pacs{74.70.Ad}{Metals; alloys and binary compounds (including
A15, Laves phases, etc.)}

\begin{document}
\maketitle

\begin{abstract}
The real $R_s(T)$ and imaginary $X_s(T)$ parts of the
surface impedance $Z_s(T)=R_s(T)+iX_s(T)$ in
polycrystalline MgB$_2$ samples of different density with
the critical temperature $T_c\approx 38$~K are measured at
the frequency of 9.4~GHz and in the temperature range $5\le
T<200$~K. The normal skin-effect condition $R_s(T)=X_s(T)$
at $T\ge T_c$ holds only for the samples of the highest
density with roughness sizes not more than 0.1~$\mu$m. For
such samples extrapolation $T\to 0$ of the {\it linear} at
$T<T_c/2$ temperature dependences
$\lambda_L(T)=X_s(T)/\omega\mu_0$ and $R_s(T)$ results in
values of the London penetration depth $\lambda_L(0)\approx
600$~\AA~and residual surface resistance $R_{res}\approx
0.8$~m$\Omega$. In the entire temperature range the
dependences $R_s(T)$ and $X_s(T)$ are well described by the
modified two-fluid model.
\end{abstract}

The inconclusive situation with the mechanism of
superconductivity in MgB$_2$ \cite{Nag} is mainly due to
the lack of consensus on many important physical quantities
in this material. For example, the energy gap ratio
$2\Delta/kT_c$ ranging from 1 to 5 has been reported, which
raises the possibility of an anisotropic energy gap or a
multiple gap. Indeed, recent tunneling \cite{Giu} and point
contact \cite{Sza} experiments showed two distinct
conductance peaks of different magnitude, and to describe
the specific heat data \cite{Wan} the authors involved
either two gaps or an anisotropic gap. At the same time a
number of high-frequency measurements in MgB$_2$ films
found a small single gap (the gap ratio is well below the
weak coupling value) if the data was fitted by BCS model
with isotropic $s$-wave order parameter
\cite{Jun,Kai,Kle,Lam}. In particular, the exponential
temperature dependence of the magnetic field penetration
depth at $T\ll T_c$ gave occasion to such fitting
\cite{Kle,Lam}. However, in very dense polycrystalline
MgB$_2$ samples the linear $\Delta\lambda(T)\propto T$
behavior was observed \cite{Li,Dul}. Other measurements in
MgB$_2$ powders and ceramics provided rather contradictory
results: $\Delta\lambda(T)$ dependences were proportional
to $T$ (Ref.\cite{Zhu2}), $T^2$ (Ref.\cite{Pan}), $T^{2.7}$
(Ref.\cite {Che}) and exp$(-\Delta/T)/T^{0.5}$
(Ref.\cite{Man}). The estimated values of $\lambda(0)$
varied from as low as 85~nm in Ref.\cite{Pan} up to 300~nm
in Ref.\cite{Kai}. Apparently, the data spread indicates
that the measured parameters of MgB$_2$ depend on the
sample quality and growth technique. Nevertheless, it is
generally accepted that in the $ab$-plane of MgB$_2$ the
coherence length $\xi_0$ does not exceed 100~\AA~and,
hence, MgB$_2$ samples characterized by the value of
electron relaxation rate $1/\tau(T_c)<10^{13}$~s$^{-1}$
(the mean free path $l>500$~\AA) should be classified as
London ($\xi_0 \ll \lambda$) pure ($l\gg\xi_0$)
superconductors.

Microwave measurements of the surface impedance temperature
dependences $Z_s(T)=R_s(T)+iX_s(T)$ provide accurate
determination of applied and fundamental parameters of the
superconducting state (residual surface resistance
$R_{res}\equiv R_s(T\to 0)$, superconducting gap and its symmetry,
penetration depth) and normal state (normal or anomalous skin-effect,
resistivity $\rho$, relaxation rate). At the frequency of 10~GHz
the skin depth $\delta=\sqrt{2\rho/\omega\mu_0}> 1$~$\mu$m is much
greater than $l$ in MgB$_2$ and, therefore, at $T\ge T_c$ the
criterion of the normal skin-effect should apply:
\begin{equation}
R_s(T)=X_s(T)=\rho(T)/\delta(T)=\sqrt{\omega\mu_0\rho(T)/2}\,.
\end{equation}

Eq.~(1) follows from the general relation between the impedance and
conductivity ($\sigma=\sigma_1-i\sigma_2$) components in London
superconductors:
\begin{equation}
\label{RS}
R_s=\sqrt {{\frac{{\omega\mu_0(\varphi ^{1/2}-1)} }{{2\sigma _2\varphi }}}},
\quad
X_s=\sqrt {{\frac{{\omega\mu_0(\varphi ^{1/2}+1)} }{{2\sigma _2\varphi }}}},
\end{equation}
where $\varphi = 1+(\sigma _1/\sigma _2)^2$. It is obvious that
$R_s\le X_s$. As follows from Eq.~(2) in the immediate vicinity of the
transition temperature there is a very narrow peak in the curve $X_s(T)$.
When $(\sigma_1/\sigma_2)^2\ll 1$, which is valid at temperatures not
very close to $T_c$, where the magnetic field penetration depth
$\lambda=\sqrt{1/\omega\mu_0\sigma_2}$, from Eq.~(2) we get
$R_s\approx\omega^2\mu_0^2\sigma_1\lambda^3/2$ and
$X_s\approx\omega\mu_0\lambda$.
In pure London superconductor the value of $\lambda$ is the London
penetration depth which equals $\lambda_L(0)=m/\mu_0ne^2$ at $T=0$.

In the centimeter wavelength band the surface resistance $R_s(T)$
was measured in MgB$_2$ wire \cite{Hakim}, pellets \cite{Dul,Hakim},
separated grains \cite{Zhu}, and films \cite{Kle,Zhu2,Lee}.
The value of $R_{res}$ characterizing the sample quality was proved
to be highly dependent on the sample preparation and processing.
The transformed ($\propto\omega^2$) to the same frequency 10~GHz
values of $R_{res}$ varied from 0.7 to 5~m$\Omega$ in different
MgB$_2$ bulk samples \cite{Kle,Zhu2,Hakim,Zhu,Lee} except the wire
\cite{Hakim} where the lower-range $R_{res}\approx 0.1$~m$\Omega$ is
found. However, in all previous publications devoted to microwave
investigations of MgB$_2$ there had been no data on the $X_s(T)$
in absolute units which means that Eq.~(1) applicability was not
verified nor was the value of $\lambda(0)=X_s(0)/\omega\mu_0$
determined directly from microwave measurements of MgB$_2$.

In this Letter, we present precise measurements of the real and
imaginary parts of the surface impedance in polycrystalline MgB$_2$
samples of different porosity. We found that normal skin-effect
condition $R_s(T)=X_s(T)$ at $T\ge T_c$ holds only for the samples
of the highest (close to theoretical) density. For such samples
the electrodynamic parameters of the normal and superconducting
state of MgB$_2$ are obtained.

The samples investigated were synthesized in situ from amorphous
boron powder and lump metal magnesium, both with purity better
than 99.96\%. Two types of different MgB$_2$ polycrystalline bulk
samples were investigated. The samples of the first type were
synthesized at 1100$^\circ$C with subsequent rapid cooling, and
the samples of the second type were obtained after heating MgB$_2$
up to 1400$^\circ$C and then keeping this temperature for an hour
(for details see Ref.\cite{Kol}). Using a diamond circular saw
we cut small thin plates ($\sim 1\times 1\times 0.1$~mm$^3$)
from the obtained compact ceramic cylinders and polished their
surfaces using fine ($\sim 0.1$~$\mu$m) diamond powder diluted by
the high purity benzine (ethanol initially used for this purpose
resulted in large microwave absorption of the sample). After
polishing and cleaning, the samples were being heated at
200$^\circ$C for 4 hours in high vacuum. The average grain size of
about $\sim 20$~$\mu$m and density of 2.52~g/cm$^3$ were obtained
for the samples of the first type and the corresponding values of
$\sim 40$~$\mu$m and 2.23~g/cm$^3$ for the samples of the second
type. This density drop was due to Mg-evaporation resulting in
numerous small voids (5-30 microns) in the ceramics. This can be
seen in Fig.~1,
\begin{figure}
\onefigure[width=14cm]{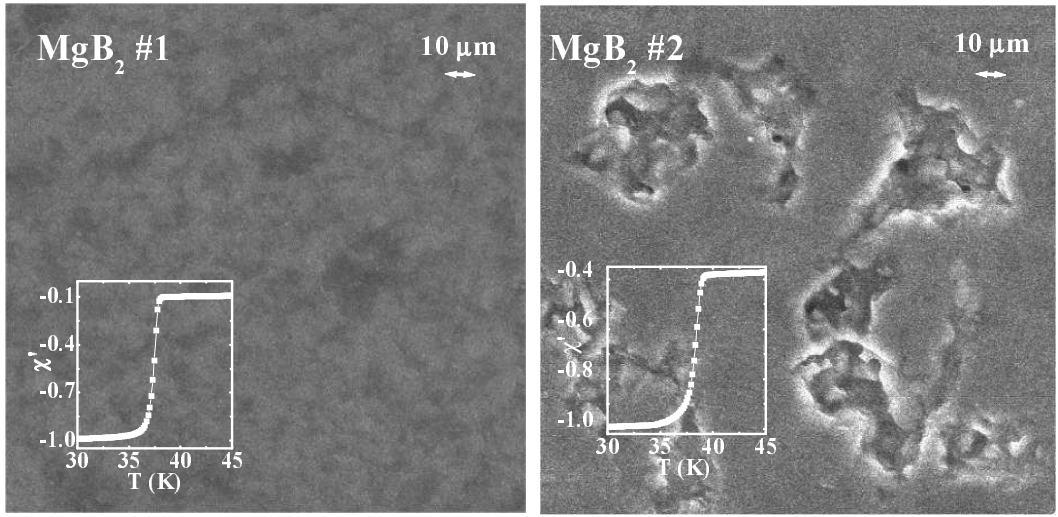} \caption{SEM micrographs of the
sample \#1 (a) and \#2 (b) in the same scale. The insets show the
$\chi'(T)$ curves of these samples in the vicinity of the
superconducting transition.} \label{f1}
\end{figure}
where scanning electron micrographs (second electron emission
mode) of the samples \#1 and \#2 are shown. Sample \#1 belongs to
the first type (dimensions: $1.34\times 1.10\times 0.03$~mm$^3$)
and sample \#2 is of the second type ($1.57\times 1.56\times
0.22$~mm$^3$). Unlike sample \#2 the surface of the densest sample
\#1 looks homogeneous, at least the dimensions of possible
irregularities are less than scanning resolution of 0.1~$\mu$m.

Preliminary testing of MgB$_2$ samples was carried out with the aid
of the temperature measurements of their dynamic susceptibility
$\chi(T)=\chi'(T)-i\chi''(T)$ performed by a four-coil scheme at
100~kHz. The insets in Fig.~1 show very similar temperature
dependences of the shielding $\chi'(T)$ of the magnetic field in both
samples. If the penetration depth
$\lambda(T)=\lambda(0)+\Delta\lambda(T)$ is much smaller
than the characteristic size $a$ (which is the size of either the
sample or grains in it), then at low temperatures $T<T_c$ the
dependence $\chi'(T)$ corresponds to the $\lambda(T)$ variation:
$\chi'(T)=\chi'_0(1-\Delta\lambda(T)/a)$.

For the measurements at 9.4~GHz we applied the "hot-finger"
technique which had been used to characterise the microwave
properties of small crystals of high-$T_c$ and conventional
superconductors \cite{Tru1}. The temperature dependences of
the Q-factor, $Q^{-1}\propto R_s$, and the frequency shift
$\Delta f\propto\Delta X_s$ of the cavity with the sample
inside are measured simultaneously. To determine $R_s(T)$
and $X_s(T)$ in absolute units one needs to know the value
of the sample geometrical factor $\Gamma$ and the constant
$X_0=X_s(T)-\Delta X_s(T)$. The parameter $\Gamma$ can be
determined both empirically and theoretically and its value
depends on the shape and dimensions of the sample and its
position in the cavity with respect to the microwave field
$\vect{H}_{\omega}$ \cite{Tru1}. The constant $X_0$ can be
derived from the condition of equality between the real and
imaginary parts of the normal state surface impedance.
Temperature dependences of the surface resistance (open
squares) and reactance (open circles) in both samples are
shown in Fig.~2.
\begin{figure}
\onefigure[width=14cm]{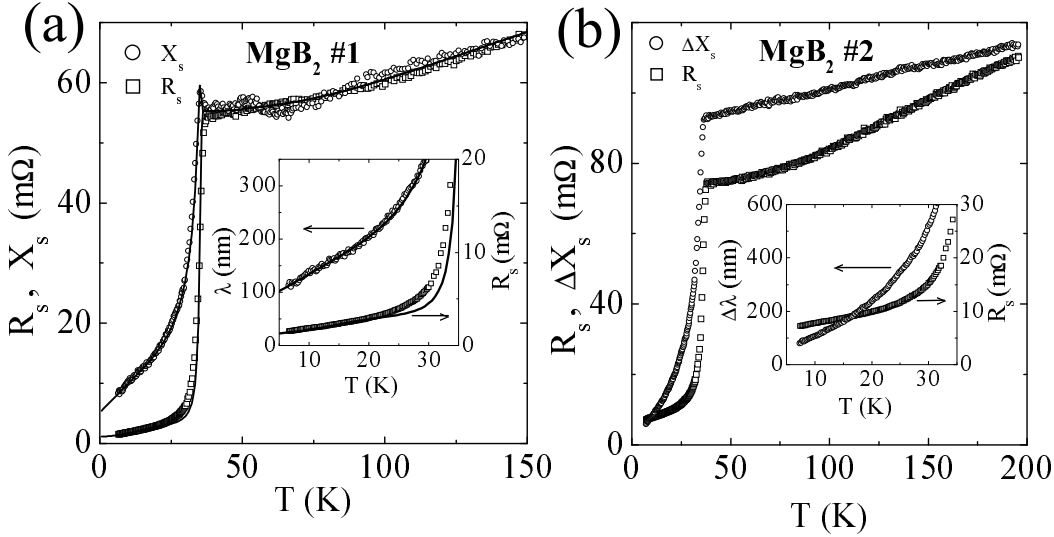} \caption{Plots of the surface
resistance $R_s(T)$ and reactance $X_s(T)$ at 9.4~GHz in MgB$_2$
samples \#1 (a) and \#2 (b). The insets show the low temperature
sections of $R_s(T)$ and London penetration depth
$\lambda_L(T)=X_s(T)/\omega\mu_0$. The solid curves in Fig.~2(a)
show calculations based on the modified two-fluid model.}
\label{f2}
\end{figure}
$Z_s(T)$ curves measured in the field $\vect{H}_{\omega}$
perpendicular and parallel to the biggest faces of the
sample were identical. The resisitivity
$\rho(T_c)=2R_s^2(T_c)/\mu_0\omega$ obtained from the
surface resistance value $R_s(T_c)$ is equal to
8~$\mu\Omega$cm and 15~$\mu\Omega$cm for samples \#1 and
\#2 respectively. The measured value of $T_c\approx 38$~K
and the residual resistivity ratio
RRR=$\rho(300$K$)/\rho(T_c)\approx 3$ for both samples
corresponds to the nominal composition Mg:B=1:2
(Refs.\cite{Chen,Ind}).

Essential distinction in samples \#1 and \#2 appears in
their $\Delta X_s(T)$ dependences. Whereas the sample \#1
parameters are in accordance with Eqs.~(1), (2):
$R_s(T)=X_s(T)$ at $T\ge T_c$, and $\Delta X_s(T)<\Delta
R_s(T)$ at $T<T_c$, the sample \#2 reactance variation
$\Delta X_s(T)$ is considerably larger than that of the
resistance $\Delta R_s(T)$ in the entire temperature range.
Unusally large change $\Delta X_s(T)>\Delta R_s(T)$ was
previously reported in the microwave $ab$-response of
Tl$_2$Ba$_2$CuO$_{6+\delta}$ single crystals
\cite{Waldr,Tru2,Kus}. As a possible explanation in
Ref.\cite{Tru2} it was proposed to allow for the shielding
effect of the microwave field by roughnesses (cleavage
plane traces) of Tl$_2$Ba$_2$CuO$_{6+\delta}$ crystal
surface. If the penetration depth is much less than
roughness sizes, both components of the effective surface
impedance measured will increase in comparison with their
values for a flat surface by the same factor equal to the
ratio of the real and flat surface areas. If the roughness
sizes are comparable to the penetration depth the situation
may occur when $\vect{H}_{\omega}$ is slightly distorted by
the roughness, whereas the high-frequency current caused by
the field decays noticably \cite{Mende}. In this case the
effective reactance ($\sim\omega\mu_0\int{H^2_{\omega}dV}$)
will exceed the sample surface resistance
($\sim\int{j^*_{\omega}E_{\omega}dV}$). It is likely that
this is the case in MgB$_2$ sample \#2 whose pores'
dimensions proved to be comparable with its skin depth.

Another non-trivial feature of $Z_s(T)$ curves in Fig.~2 is linear
temperature dependence of both impedance components at $T<T_c/2$
(see the insertions in Fig.~2). Fig.~3 demonstrates an agreement of
the temperature dependences $\Delta\lambda(T)\propto T$
measured at the frequencies 9.4~GHz and 100~kHz.
\begin{figure}
\onefigure[width=7cm]{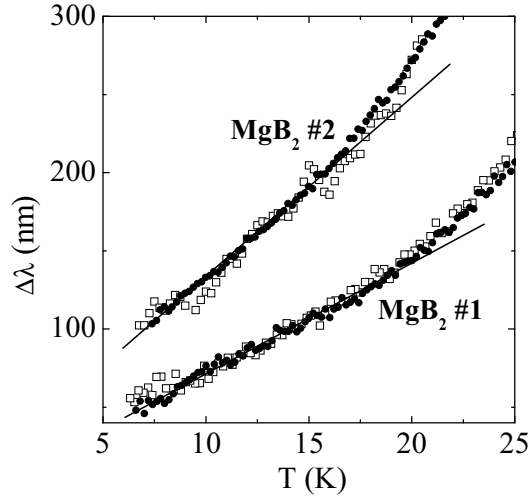} \caption{Comparison of
$\Delta\lambda (T)$ data at 9.4~GHz (black circles) and at 100~kHz
(open squares) in MgB$_2$ samples \#1 and \#2. The upper graph is
shifted 20~nm up for clearness.} \label{f3}
\end{figure}
Characteristic sizes $a=\Delta\lambda(T)/\Delta\chi'(T)$,
determined for the samples \#1 and \#2 with $\chi'_0=-1$, proved
to be equal to the average grain sizes respectively. In the sample
\#1 at $T<T_c/2$ the slopes of $R_s(T)$ and $X_s(T)$ curves are
roughly half of that of the sample \#2. In particular, the value
of $d\lambda/dT\approx 70$~\AA/K at $T\ll T_c$ for the sample \#1
so we can derive the value of $\lambda_L(0)=600\pm 100$~\AA.
Extrapolation $R_s(T\to 0)$ gives $R_{res}\approx 0.8$~m$\Omega$
in this sample. The strong slope of the linear sections of
$\Delta\lambda(T)$ curves in very dense polycrystalline MgB$_2$
samples was also found in Refs.\cite{Li,Dul}. The above value of
$\lambda_L(0)$ is smaller then the one found previously in
microwave investigations of MgB$_2$. One should notice that
elsewhere the values of $\lambda(0)$ were obtained as a result of
fitting of the measured $\Delta\lambda(T)$ curve by model
dependence of $\lambda(T)$, whereas we obtain the $\lambda_L(0)$
value directly from the measurements of $X_s(0)$.

Interestingly, $R_s(T)$ and $X_s(T)$ curves in Fig.~2(a) are very
similar to those measured in electron high-$T_c$ crystal
Ba$_{0.6}$K$_{0.4}$BiO$_3$ \cite{Tru3}. Moreover, the similarity of
the optical constants in
MgB$_2$ and Ba$_{0.6}$K$_{0.4}$BiO$_3$ was pointed out in Ref.\cite{Tu}.
So far there is no microscopic theory explaining such a strong slope of
the linear $Z_s(T)$ temperature dependence at $T$ up to $T_c/2$ for
these crystals. As early as in the old Ref.~\cite{Tru3} a modified
two-fluid model (MTFM) was suggested and then developed in
Refs.~\cite{Fink} to describe $Z_s(T)$ in
all optimum doped high-$T_c$ single crystals. This phenomenological model
has two essential features different from the well-known Gorter-Casimir
(GC) model \cite{Gor}. The former is the unique density of superconducting
electrons which gives rise to a linear temperature dependence of
the penetration depth at low temperatures, and the latter is the
introduction of the temperature dependence of the quasiparticle
relaxation time $\tau(T)$ described by Gr\"uneisen formula
(electron--phonon interaction) with retaining the temperature-independent
impurity relaxation time $\tau(0)$, which is present in GC model:
\begin{equation}
\begin{array} {c}
{1 \over {\tau}}={1 \over {\tau (0)}}\left[1+{{t^5{\cal J}_5(\kappa/t)
/{\cal J}_5(\kappa )} \over {\beta }}\right], \\
{{\cal J}_5(\kappa/t)=\int\limits_0^{\kappa /t}{z^5e^zdz
\over {(e^{z}-1)^2}}},
\end{array}
\end{equation}
where $t\equiv T/T_c$, $\kappa =\Theta /T_c$ ($\Theta$ is
the Debye temperature), and $\beta$ is a numerical
parameter. From Eq.~(3) we have
$\beta=\tau(T_c)/[\tau(0)-\tau(T_c)]$; $\beta \approx
\tau(T_c)/\tau(0) \ll 1$ if $\tau(0)\gg\tau(T_c)$, and
$\beta\gg 1$ if $\tau(0)\approx\tau(T_c)$. The second
summand in the brackets in Eq.~(3) is proportional to $T^5$
at $T<\Theta/10$ and to $T$ at $T>\Theta/5$. Solid line at
$T\ge T_c$ in Fig.~2(a) shows the calculation result of
$R_s(T)=X_s(T)$ from Eq.~(1) with parameters $\beta=350$
and $\kappa=15$ ($\Theta\sim 600$~K) in Eq.~(3) for
$1/\tau(T)\propto\rho(T)$. In the framework of MTFM the
value of $\tau(T_c)=X_s^2(0)/2\omega R_s^2(T_c)\approx
0.6\cdot 10^{-13}$~s, $\tau(0)=dR_s/\omega dX_s|_{T\to
0}\approx 6.6\cdot 10^{-13}$~s (obtained using the measured
slopes ${dR_s}/{dT}$ and ${dX_s}/{dT}$ at $T\ll T_c$), and,
hence, $\beta=0.1$ in Eq.~(3) for the superconducting state
of sample \#1. Solid lines at $T\le T_c$ in Fig.~2(a) show
the components $R_s(T)$ and $X_s(T)$ calculated from
Eq.~(2). Thus, $Z_s(T)$ dependences measured in MgB$_2$ are
well described in the framework of MTFM in the entire
temperature range with the same free parameter $\kappa=15$
in Eq.~(3), but with substantially different $\beta$ in the
normal and superconducting state. As in the case with
high-$T_c$ crystals, in MgB$_2$ the relaxation rate of
normal carriers decreases rapidly ($\propto T^5$) with
decrease of temperature lower than $T_c$. In principle,
there is a possibility to describe the linear temperature
dependences of $R_s(T)$ and $X_s(T)$ at low $T$ in terms of
weak links model \cite{Halbritter}. However, we did not
observe any manifestation of intergranular weak links in
our measurements of ac-susceptibility in a weak magnetic
field. Moreover, magnetization \cite{Larb}, transport
\cite{Kim} and non resonant microwave absorption
\cite{Joshi} measurements show that MgB$_2$ does not
exhibit weak-link electromagnetic behavior.

In conclusion, we state that the measurement results of the microwave
surface impedance components in polycrystalline MgB$_2$ samples
depend on their quality and preparation technique. The normal skin-effect
criterion $R_s(T)=X_s(T)$ at $T\ge T_c$ is met only in very dense
homogenious samples of the highest quality characterized by the
lower-range values of residual losses, superconducting transition width,
and resistivity in the normal state. For these samples the values of
$\tau(T_c)\approx 10^{-13}$~s, $\lambda_L(0)\approx 600$~\AA~and
$R_s(0)\approx 0.8$~m$\Omega$ are obtained directly from $R_s(T)$
and $X_s(T)$ measurements in absolute units. The linear dependences
$\Delta\lambda (T)\propto T$ and $R_s(T)\propto T$ at $T<T_c/2$ are
observed. The curves $\Delta\lambda (T)$ measured at 9.4~GHz coincide
with ac-susceptibility measuremens at five orders smaller frequency.
$R_s(T)$ and $X_s(T)$ curves in the normal and superconducting state
are well described in the framework of the phenomenological approach
based on the usage of Eq.~(3) for quasiparticle scattering on
impurities and phonons.

This research has been supported by RFBR grant 00-02-17053, DFG-RFBR
grant 00-02-04021, Scientific Council on Superconductivity (project 96060),
and University of Palermo (grant Coll. Int. Li Vigni).


\begin{thebibliography}{0}

\bibitem{Nag}
\Name{Nagamatsu~J., Nakagawa~N., Muranaka~T., Zenitani~Y. \and
Akimitsu~J.} \REVIEW{Nature}{63}{2001}{410}.

\bibitem{Giu}
\Name{ Giubileo~F., Roditchev~D., Sacks~W., Lamy~R., Thanh~D.X.,
Klein~J., Miraglia~S., Fruchart~D., Marcus~J. \and Monod~Ph.}
\REVIEW{Phys.Rev.Lett.}{87}{2001}{177008}.

\bibitem{Sza}
\Name{Szabo~P., Samuely~P., Kacmarcik~J., Klein~Th., Marcus~J.,
Fruchart~D., Miraglia~S., Marcenat~C. \and Jansen~A.G.M.}
\REVIEW{Phys.Rev.Lett}{87}{2001}{137005}.

\bibitem{Wan}
\Name{Junod~A., Wang~Y., Bouquet~F., Toulemonde~P.}
cond-mat/0106394; \Name{Bouquet~F., Fisher~R.A., Phillips~N.E.,
Hinks~D.G. \and Jorgensen~J.D. }
\REVIEW{Phys.Rev.Lett.}{87}{2001}{047001}.

\bibitem{Jun}
\Name{Jung~J.H., Kim~K.W., Lee~H.J., Kim~M.W., Noh~T.W.,
Kang~W.N., Kim Hyeong-Jin, Choi Eun-Mi, Jung~C.U. \and Lee
Sung-Ik} cond-mat/0105180.

\bibitem{Kai}
\Name{Kaindl~R.A., Carnahan~M.A., Orenstein~J., Chemla~D.S.,
Christen~H.M., Zhai~H-Y., Paranthaman~M. \and Lowndes~D.H.}
cond-mat/0106342.

\bibitem{Kle}
\Name{Klein~N., Jin~B.B., Schubert~J., Schuster~M., Yi~H.R.,
Pimenov~A., Loidl~A. \and Krasnosvobodtsev~S.I.} cond-mat/0107259.

\bibitem{Lam}
\Name{Lamura~G., Gennaro~E.~Di, Salluzzo~M., Andreone~A.,
Cochec~J.~Le, Gauzzi~A., Cantoni~C., Paranthaman~M.,
Christen~D.~K., Christen~H.~M., Giunchi~G. \and Ceresara~S.}
preprint.

\bibitem{Li}
\Name{Li~S.L., Wen~H.H., Zhao~Z.W., Ni~Y.M., Ren~Z.A., Che~G.C.,
Yang~H.P., Liu~Z.Y. \and
Zhao~Z.X.}\REVIEW{Phys.Rev.B}{64}{2001}{094522}.

\bibitem{Dul}
\Name{Dulcic~A., Paar~D., Pozek~M., Williams~G.V.M., Kramer~S.,
Jung~C.U., Park Min-Seok \and Lee Sung-Ik} cond-mat/0108071.

\bibitem{Zhu2}
\Name{Zhukov~A.A., Cohen~L.F., Purnell~A., Bugoslavsky~Y.,
Berenov~A., MacManus-Driscoll~J.L., Zhai~H.Y., Christen Hans~M.,
Paranthaman Mariappan~P., Lowndes Douglas~H., Jo~M.A.,
Blamire~M.C., Hao Ling \and Gallop~J.} cond-mat/0107240.

\bibitem{Pan}
\Name{Panagopoulos~C., Rianford~B.D., Xiang~T., Scott~C.A.,
Kambara~M. \and Inoue~I.H.} \REVIEW{Phys.Rev.B}{64}{2001}{094514}.

\bibitem{Che}
\Name{Chen~X.H., Xue~Y.Y., Meng~R.L. \and Chu~C.W.}
cond-mat/0103029.

\bibitem{Man}
\Name{Manzano~F. \and Carrington~A.} cond-mat/0106166.

\bibitem{Hakim}
\Name{Hakim~N., Parimi~P.V., Kusko~C., Sridhar~S., Canfield~P.C.,
Bud'ko~S.L. \and Finnemore~D.K.} \REVIEW{Appl. Phys.
Lett.}{17}{2001}{4160}.

\bibitem{Zhu}
\Name{Zhukov~A.A., Yates~K., Perkins~G.K., Bugoslavsky~Y.,
Polichetti~M., Berenov~A., Driscoll~J., Caplin~A.D., Cohen~L.F.,
Hao~L. \and Gallop~J.}
\REVIEW{Supercond.Sci.Technol.}{14}{2001}{L13}.

\bibitem{Lee}
\Name{Lee~S.Y., Lee~J.H., Lee Jung~H., Ryu~J.S., Lim~J.,
Moon~S.H., Lee~H.N., Kim~H.G. \and Oh~B.} cond-mat/0105327.

\bibitem{Kol}
\Name{Kolesnikov~N.N. \and Kulakov~M.P.} \REVIEW{Physica C}
{363}{2001}{166}.

\bibitem{Tru1}
\Name{Trunin~M.R.} \REVIEW{J. Supercond.}{11}{1998}{381}.

\bibitem{Chen}
\Name{Chen~X.H., Wang~Y.S., Xue~Y.Y., Meng~R.L., Wang~Y.Q. \and
Chu~C.W.} cond-mat/0107154.

\bibitem{Ind}
\Name{Indenbom~M.V., Uspenskaya~L.S., Kulakov~M.P., Bdikin~I.K.,
\and Zver'kov~S.A.} \REVIEW{JETP Lett.}{74}{2001}{274}.

\bibitem{Waldr}
\Name{Waldram~J.R., Broun~D.M., Morgan~D.C., Ormeno~R. \and
Porch~A.} \REVIEW{Phys. Rev. B}{59}{1999}{1528}.

\bibitem{Tru2}
\Name{Trunin~M.R.} \REVIEW{JETP Lett.}{72}{2000}{583}.

\bibitem{Kus}
\Name{Kusko~C., Zhai~Z., Hakim~N., Markievicz~R.S., Sridhar~S.,
Colson~D., Viallet-Guillen~V., Nefyodov~Yu.A., Trunin~M.R.,
Kolesnikov~N.N., Maignan~A., \and Erb~A.} submitted to {\it Phys.
Rev. Lett.}

\bibitem{Mende}
\Name{Mende~F.F. \and Spitsyn~A.A.}
  \Book{Surface impedance of superconductors}
  \Publ{Naukova dumka, Kiev}
  \Year{1985}.

\bibitem{Tru3}
\Name{Trunin~M.R., Zhukov~A.A., Tsydynzhapov~G.E., Sokolov~A.T.,
Klinkova~L.A. \and Barkovskii~N.V.} \REVIEW{JETP
Lett.}{64}{1996}{832}.

\bibitem{Tu}
\Name{Tu~J.J., Carr~G.L., Perebeinos~V., Homes~C.C., Strongin~M.,
Allen~P.B., Kang~W.N., Choi~E.M., Kim~H.J. \and Lee Sung-Ik}
cond-mat/0107349.

\bibitem{Fink}
\Name{Fink~H.J.} \REVIEW{Phys. Rev. B}{58}{1998}{9415}, {\bf 61}
(2000) 6346; \Name{Trunin~M.R., Nefyodov~Yu.A. \and Fink~H.J.}
\REVIEW{JETP}{91}{2000}{801}; \Name{Fink~H.J. \and Trunin~M.R.}
\REVIEW{Phys. Rev. B}{62}{2000}{3046}.

\bibitem{Gor}
\Name{Gorter~C.S. \and Casimir~H.}
\REVIEW{Phys.Z.}{35}{1934}{963}.

\bibitem{Halbritter}
\Name{Halbritter~J.} \REVIEW{J. Supercond.}{8}{1995}{691}.

\bibitem{Larb}
\Name{Larbalestier~D.C., Rikel~M.O., Cooley~L.D.,
Polyanskii~A.A., Jiang~J.Y., Patnaik~S., Cai~X.Y.,
Feldmann~D.M., Gurevich~A., Squitieri~A.A., Naus~M.T.,
Eom~C.B., Hellstrom~E.E., Cava~R.J., Regan~K.A., Rogado~N.,
Hayward~M.A., He~T., Slusky~J.S., Khalifah~P., Inumaru~K.,
Haas~M.} \REVIEW{Nature}{410}{2001}{186}.

\bibitem{Kim}
\Name{Kang~W.N., Kim Kijoon H.P., Kim Hyeong-Jin, Choi
Eun-Mi, Park Min-Seok, Kim Mun-Seog, Du Zhonglian, Jung
Chang Uk, Kim Kyung Hee, Lee Sung-Ik, Mun Mi-Ock}
cond-mat/0103161.

\bibitem{Joshi}
\Name{Joshi~Janhavi P., Sarangi~Subhasis, Sood~A.K.,
Bhat~S.V., Pal~Dilip} cond-mat/0103369.

\end{thebibliography}
\end{document}